\documentstyle[amssymb,11pt]{article}
\renewcommand\baselinestretch{1.3}
\input{tcilatex}
\begin{document}

\title{{\bf Firms Growth Dynamics, Competition and Power Law Scaling}}
\author{Hari M. Gupta and Jos\'{e} R. Campanha \\
%EndAName
Departamento de F\'{\i }sica- IGCE \\
Universidade Estadual Paulista (UNESP)\\
C.P. 178, Rio Claro 13500-970, SP, Brazil\\
}
\date{}
\maketitle

\begin{abstract}
{\bf We study the growth dynamics of the size of manufacturing firms
considering competition and normal distribution of competency. We start with
the fact that all components of the system struggle with each other for
growth as happened in real competitive bussiness world. The detailed
quantitative agreement of the theory with empirical results of firms growth
based on a large economic database spanning over 20 years is good .Further
we find that this basic law of competition leads approximately a power law
scaling over a wide range of parameters. The empirical datas are in
accordance with present theory rather than a simple power law. \ \newline
}

\ \newline
\end{abstract}

%\begin{titlepage}
%independiza o estilo da capa
\renewcommand{\baselinestretch}{1} %comando para espaco simples
%\renewcommand{\thepage}{}
%tira o numero da primeira pagina

%\end{titlepage}

\newpage

Recently, study of natural i.e. complex physical, biological and social
systems has become a new area of physical investigations. In complex systems
a large number of elementary interactions are taking place at the same time
for large number of components and thereby make it difficult to make an
exact analysis. Thus statistical methods are normally used to study these
systems. Amaral et. al. [1], discussed firms growth dynamics based on the
interaction between different units of a complex system, which is a normal
practice in physics for simpler systems. However, we feel that in real
business firms, this procedure is not adequate because a large number of
elementary interactions are taking place at the same time including many
with outside elements. In biological and social sciences, growth and
evolution processes are normally explained qualitatively through
competition. We feel that in describing complex systems, competition is more
relevant parameter compared to internal interactions particularly in social
sciences. It is therefore interesting to study quantitatively economical
complex systems in the spirit of Darwin\'{}s classical theory of biological
evolution: Survival for the fittest. This will also provide an alternative
thinking in understanding of complex systems. The present model may also be
important for some other complex physical, biological and economical systems.

We start with the fact that all components of a system struggle i.e. compete
with each other for growth, like all firms in economical world compete with
each other to grow. The same is the case between different species or groups
in biological, political, religious or social circle. The components which
are better than their competitors will grow, while other shrink till reach
to a equilibrium state where they are equal to an average competitor of
their class. We formulate our problem for the growth of the manufacturing
firms, because there is plenty of quantitative data [1-5] and the term
competition and competency is well understood. The problem is also important
due to considerable recent interest in economics in developing a richer
theory of the growth dynamics of a firm [5-10]. The present model is
interesting as it is close to real cut-throat competitive business..

In economics, what is commonly called ``theory of the firm'' is actually a
theory of a business unit [7,10]. Thus the standard model of the firm does
not yield any prediction about the distribution of the size of actual,
multi-divisional firms or their growth rates. In the present paper, we
explore the stochastic properties of the dynamics of firm growth in line
with the work done by Simon [8] and Lucas [9]. Lucas suggests that the
distribution of a firm size depends on the distribution of its managerial
ability in the economy rather than on the factors that determine the size in
the conventional theory of the firm [7,10]. As all firms are competing with
each other and political, economical, and technological factors are same for
all, we also believe that in long terms, the managerial ability of a firm is
a determining factor in its growth and the managerial ability is basically
determined through its chief executive officer (CEO). The management ability
means the person\'{}s effective ability to run a firm. It includes the
ability of a person to take (a) maximum advantages in given circumstances
(b) to correctly predict future business happenings, (c) courage to make
necessary changes and take risk and (d) maintain complete harmony in whole
administration . We obtain the statistical probability distribution of the
growth rate of firms of certain size and the fluctuation in the growth
rates, measured by the width of the distribution, scale only approximately
as a power law with firm size. Finally we discuss the results with available
data.

A business firm is a very complex system. We make the following assumptions
to have a simplified version of the whole system to study the statistical
behavior of the growth rate of the firms. However for the growth of a
particular industry in a particular segment, other factors like economical,
political and technological will also be important.

(i) We consider the sales of a firm to define its size, although other
parameters like number of employee or assets can also be used [3]. The
observed sales distributions of firms are very close to log-normal
distribution [5,6] except for very large firms. In general, we expect a
normal distribution of a variable. We therefore, define size of a firm equal
to natural logarithm of its sales $(S)$.

(ii) The management ability of people vary from person to person. Here, by
management ability we mean the person\'{}s inherent ability to run a
pre-established business after proper training. We, define quantitatively
management ability M of a person (or competitive ability in general) in
arbitrary units, equal to size of the firm under his supervision, when the
firm is in equilibrium i.e. neglecting short term fluctuations, the growth
rate is zero. In this case his management ability is sufficient only for
running the firm satisfactorily without any growth.

\medskip

\begin{equation}
M=\ln (S)|_{Equilibrium}
\end{equation}

\smallskip

(iii) When the management ability $(M)$ of a CEO is more than what is needed
to handle the firm under his control $(\ln (S)),$ the firm grows. As per
common experience of growth rate of a firm under same CEO for long time [11]
and observed probabilistic distribution of the growth rate of the firms, we
consider for mathematical simplicity that the growth rate $(\gamma )$ of the
firm is directly proportional to the relative management ability ($R$) of
its CEO, i.e.

\medskip

\begin{equation}
\gamma =\ln (\frac{S_{1}}{S})=kR
\end{equation}

\medskip

\noindent The relative management ability of a CEO, is the difference of his
management ability and management ability needed to handle the firm:

\smallskip

\begin{equation}
R=\left| M-\ln (S)\right|
\end{equation}

\smallskip

$k$ is a constant of the system. $S_{1}$ is the sales of the firm in the
following year. Under this assumption, the optimum size of a firm is
determined through management ability of its CEO $(S_{optimum}=\exp (M))$
and the firm approaches to this size [11]. In Figure 1, we have shown the
variation of annual sell of a firm with initial sell $10^{6}$ and is under a
CEO, who can handle satisfactorily a firm of annual sell $10^{8}$.

(iii) We consider that the management ability of people in general ($M_{G})$
is having a normal distribution [12] with mean $\overline{M_{G}}$ and
standard deviation $\sigma _{G}$. Thus:

\medskip

\begin{equation}
P_{G}(M)=\frac{1}{\sqrt{2\pi }\sigma _{G}}\exp -\frac{(M-\overline{M_{G}}%
)^{2}}{2\sigma _{G}^{2}}
\end{equation}

\medskip

\noindent where $P_{G}(M)$ is the probability of a person in general to have
management ability equal to $M$. We have shown it in Figure 2a.

(iv) Let us consider firms of annual sales $S_{0}.$ In general, they have
their CEO of different management ability. The firms under CEO with
management ability $M_{0}$, where $M_{0}=\ln (S_{0})$ will be in
equilibrium. We consider that this is also means of the management ability
of CEO\'{}s of these firms. Basically a person is hired as CEO of a firm
because he (a) is competent i.e. his management ability is either equal or
better than average of his group, (b) is having strong influence on the
share holders. We consider that political factors are the same for all and
all competent people have equal chance to become CEO, thus the distribution
of management ability of CEO of these firms for positive segment is in same
way as people are available.. For negative segment $(M<M_{0})$, we consider
the distribution as mirror image of the positive segment as most of the
people will be rejected on merit ground. This will give a symmetrical growth
dynamics at zero growth rate. Further the over all distributions of firm
size is invariant with time, which is more or less true. The probability of
a person to be selected as CEO must decrease with his incompetency.. For big
firms, this probability decreases more rapidly as they are more mature and
exigent. This will give truncated normal distribution with mean management
ability equal to $M_{0}$. In Figures 2b, we have drawn distribution of
relative management ability of the CEO through Equation (3).

Under these assumptions and after normalization, the probability
distribution of management ability (M) of CEO of firms of the size $S_{0}$ $%
(P(M/S_{0}))$ for $M\geqslant M_{0}$ is given by:

\medskip

\begin{equation}
P(M/S_{0})=\frac{P_{G}(M)}{2\int\limits_{M_{0}}^{\infty }P_{G}(M)dM}
\end{equation}

\medskip

The probability distribution for relative management ability $(P(R/S_{0}))$
is thus given by:

\medskip

\begin{equation}
P(R/S_{0})=\frac{P_{G}(M_{0}+R)}{2\int\limits_{0}^{\infty }P_{G}(M_{0}+R)dR}
\end{equation}

\medskip

Using equations (3), (6) and normalization condition, the probability
distribution of the growth rate of firms of size $S_{0}$ $(P(\gamma /S_{0}))$
and standard deviation of the growth rate of these firms $(\sigma (S_{0}))$
are given by:

\medskip

\begin{equation}
P(\gamma /S_{0})=\frac{P_{G}[M_{0}+\frac{\gamma }{k}]}{2k\int\limits_{0}^{%
\infty }P_{G}(M_{0}+R)dR}
\end{equation}

\smallskip

\begin{equation}
\sigma ^{2}(S_{0})=\frac{\int\limits_{0}^{\infty }P(\frac{\gamma }{S_{0}}%
)\gamma ^{2}d\gamma }{\int\limits_{0}^{\infty }P(\frac{\gamma }{S_{0}}%
)d\gamma }=k^{2}\frac{\int\limits_{0}^{\infty }P_{G}(M_{0}+R).R^{2}dR}{%
\int\limits_{0}^{\infty }P_{G}(M_{0}+R)dR}
\end{equation}

\medskip

\smallskip In Figure 2c, we have shown distribution of growth rate using
Equation (7).

We compare our theory with growth rate of all publicly traded manufacturing
companies of US. in the 1994 Compustat database with standard industrial
classification index of 2000-3999. The distribution represents all annual
growth rates observed in the 19 year period 1974-1993. We consider $k=0.6,$ $%
\overline{M_{G}}=12$ and $\sigma _{G}=1.3$ for drawing theoretical curves. $k
$ is chosen to match the magnitude of the growth rate while $\overline{M_{G}}
$ and $\sigma _{G}$ are chosen to give best fit in standard deviation $vs$.
size. The value of $\overline{M_{G}}=12.0$ signify that about $25\%$ of
people have capacity to run a pre-established manufacturing firm of annual
sell of about $5.10^{5}$ dollars, after proper training and experience,
which seems to be reasonable as this is the size of small family businesses.
In Figure 3a, we compare the probability density of the growth rate for
different initial sales while in Figure 3b, we compare the standard
deviation as a function of the initial sales. The straight line is a guide
for the eye with $slope\simeq 0.15$. We obtained all the four theoretical
curves with a single set of two chosen parameters. The agreement is
reasonably good for all the curves.

Power law scaling [13,14] is one of the main feature of complex systems and
has been speculated in many physical [15-18], chemical [19], economical
[20-23], biological [24,25], geophysical [26] and others [27,28] systems. It
has also been speculated in firm growth dynamics. The present theory
approximately leads to power law scaling. It is interesting to observe that
the general behavior of the empirical results is also close to present
theoretical prediction rather than any power law scaling (straight line
fitting). Power law scaling is only a crude approximation for these results
for a interesting range of parameters.

In the present theory, the probability of small positive or negative growth
is more than zero growth for firms of annual sell less than $10^{5}$
dollars. Normally the firms of this size either grow to become competitive
or soon get out of the market.

The standard deviation of management ability of people in general is $1.3$,
while the standard deviation of firms size for these data is $2.72$ [5].
Thus the number of competent CEO available per firm decreases very rapidly
with increase of firm size. This means that for small firms, many competent
people are available and political considerations are more important for
selection. As a firm grow, merit consideration become more and more
important. For firms of annual sell more than 10$^{8}$ dollars, it become
almost necessary to have many quasi independent executives and decisions
have to be taken collectively.. This is what happened normally. The number
of very large firms is less than what is given by log-normal distribution
[5]. Perhaps this happened because of shortage of highly competent executive
to run these firms. In view of these comments, the present theory is more
appropriate for firms with annual sell below 10$^{8}$ dollars per year. For
firms with sell larger than 10$^{8}$ dollars/year, it is necessary to
consider effective management ability of the whole management board, where
perhaps theory proposed by Amaral et. al. [1] is also useful.

In Figure 4, we draw the variation of standard deviation of the growth rate
as a function of initial sales for (a) various values of $\overline{M_{G}}$
and (b) various values of $\sigma _{G}.$ We consider $\overline{M_{G}}=12.0$%
, $\sigma _{G}=1.3,$ if it is not variable. The power law scaling behavior
is approximately maintained in all these cases. This approximate power law
scaling behavior is not very sensitive to the value of parameters under
interested experimental range. This removes the necessity of self-tuning of
parameters controlling the dynamics of the system to their critical values
as is needed in critical behavior [29-31]. Perhaps, competition and normal
distribution of competency is behind the approximate scaling behavior in
diverse fields, particularly in social and biological sciences, independent
of the microscopic details.

\newpage

\begin{center}
{\bf Figure Captions}
\end{center}

\smallskip

\noindent Figure {\bf 1}: Variation of sell of a firm with initial sell $%
10^{6}$ and is under a CEO who have management ability to run a firm of sell 
$10^{8}$.

\noindent Figure {\bf 2: }Our theoretical model {\bf (a)} probability
density distribution of management ability $(P_{G}(M)$ $vs.$ $M)$ for public
in general - a normal distribution. {\bf (b)} Probability density
distribution of relative management ability $(R=(M-\ln (S))$ for CEO of
firms of sales $S_{0}$ $(P(R/S_{0})$ $vs$. $R).$ {\bf (c)} Probability
density distribution of the growth rate for firm of sales $S_{0}$ $(P(\gamma
/S_{0})$ $vs.$ $\gamma )$.

\smallskip

\noindent Figure {\bf 3}: {\bf (a)} Probability density distribution of the
growth rate $(P(\gamma /S_{0})$ $vs.$ $\gamma )$ for all publicly traded
U.S. manufacturing firms. The solid lines are theoretical curves with $k=0.6$%
, $\overline{M_{G}}=12$ and $\sigma _{G}=1.3$. {\bf (b) }Standard deviation
of the growth rates as a function of the initial sales $(\sigma (S_{0})$ $%
vs. $ $S_{0}).$ The solid lines is the theoretical curve with same
parameters, while broken line is a guide for eye with slope$\simeq 0.15$.

\smallskip

\noindent Figure {\bf 4}: Standard deviation of the growth rate as a
function of the initial sales $(\sigma (S_{0})$ $vs$ $S_{0})$ for different
values of {\bf (a)} mean management ability of people in general $\overline{%
M_{G}}$ ({\bf b}) Standard deviation of the management ability of the people
in general $(\sigma _{G})$.

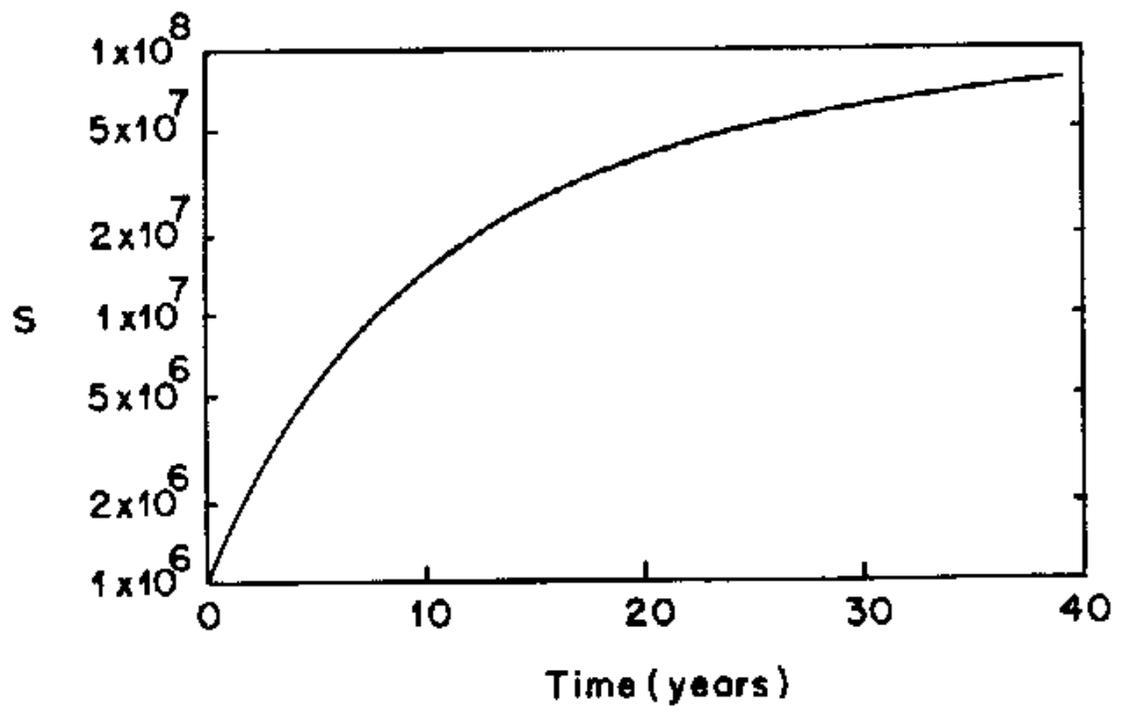
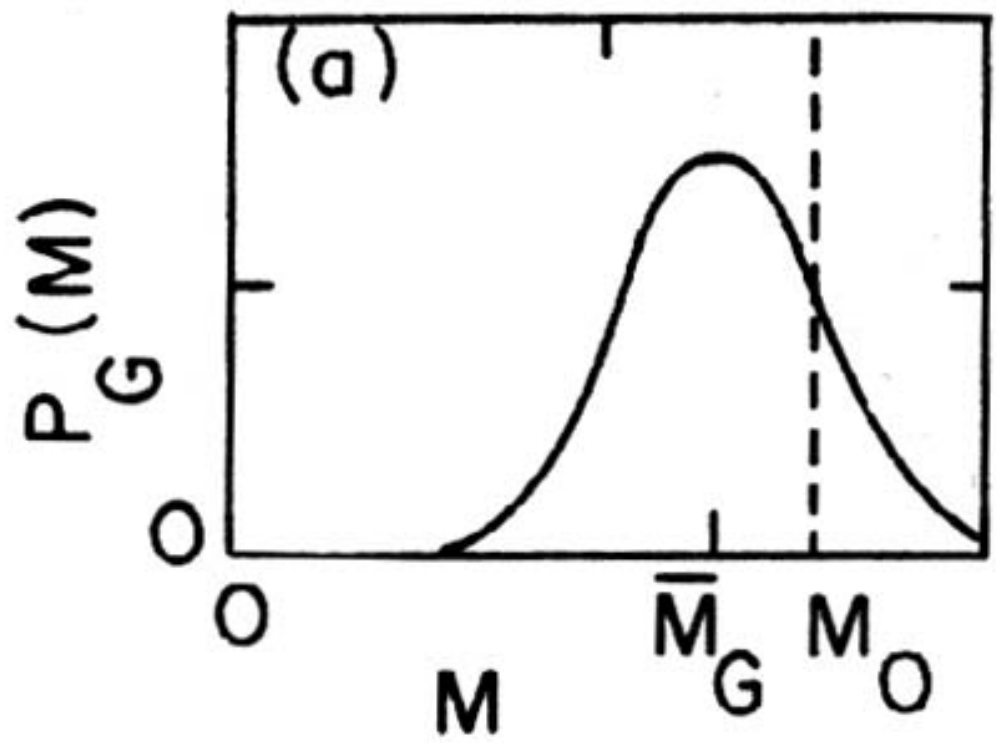
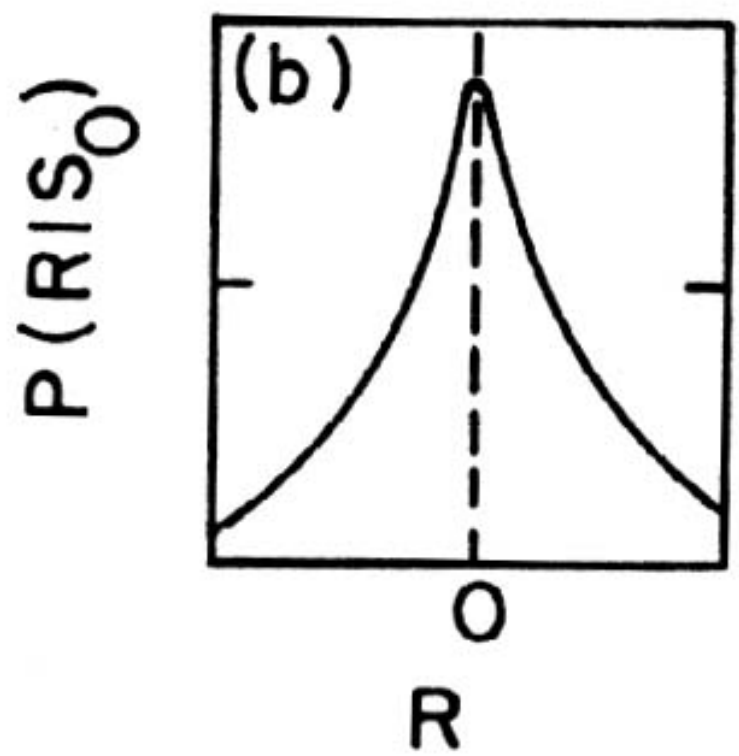
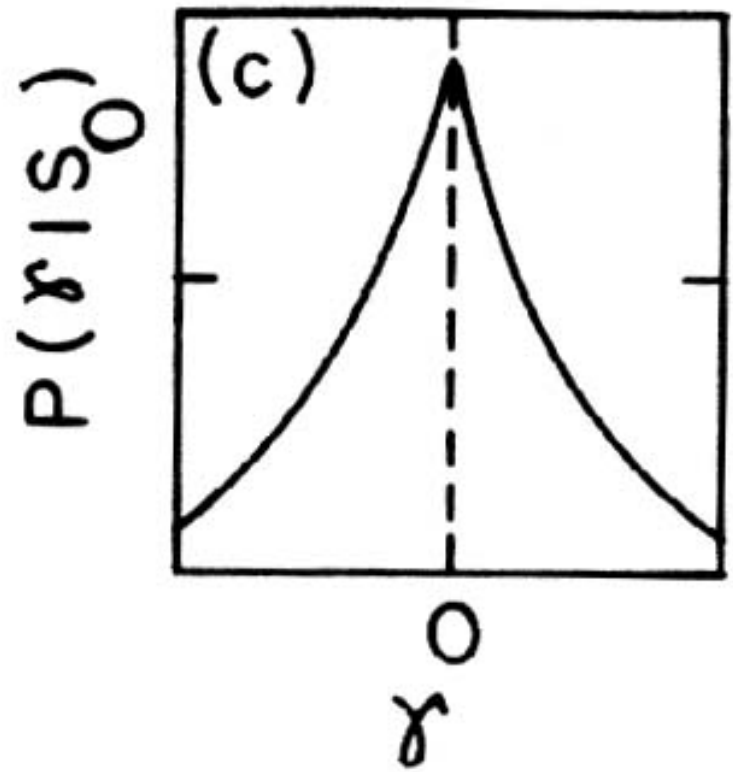
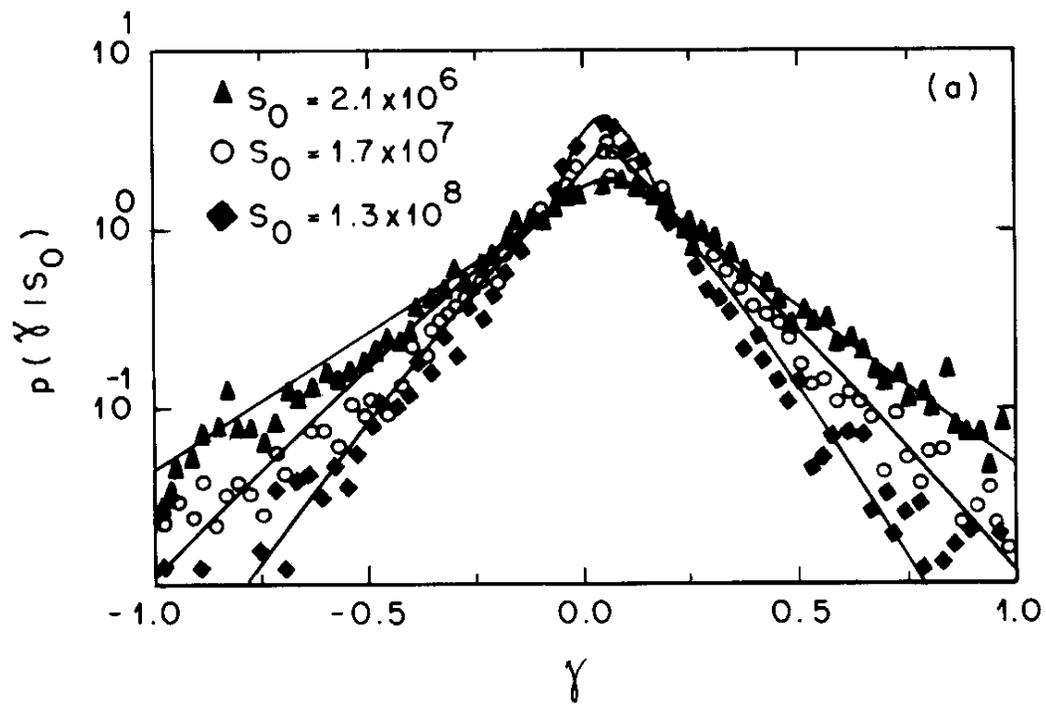
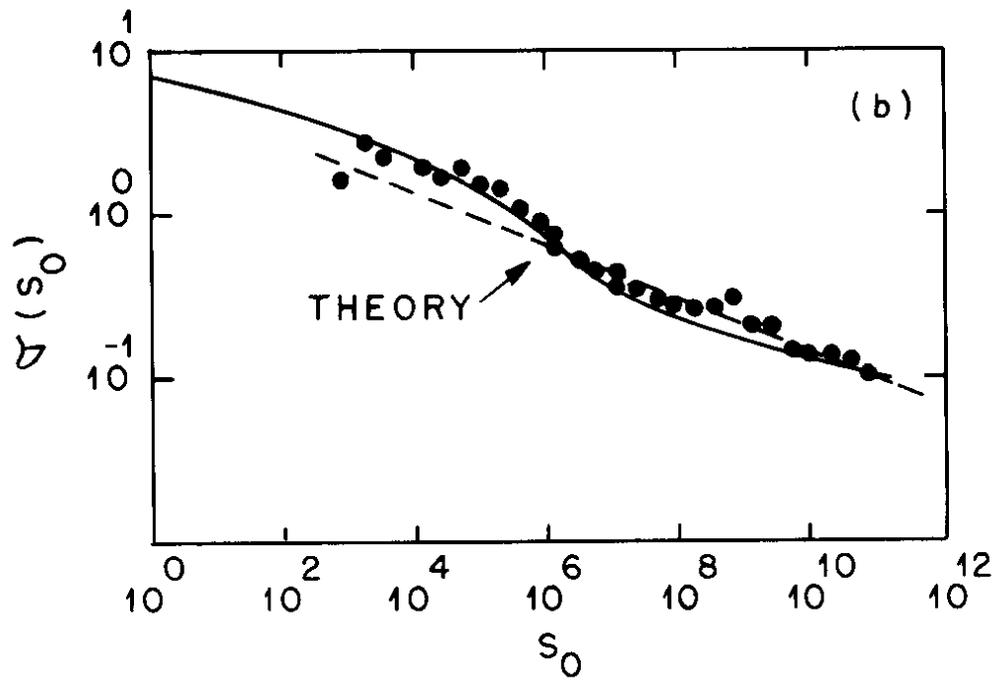
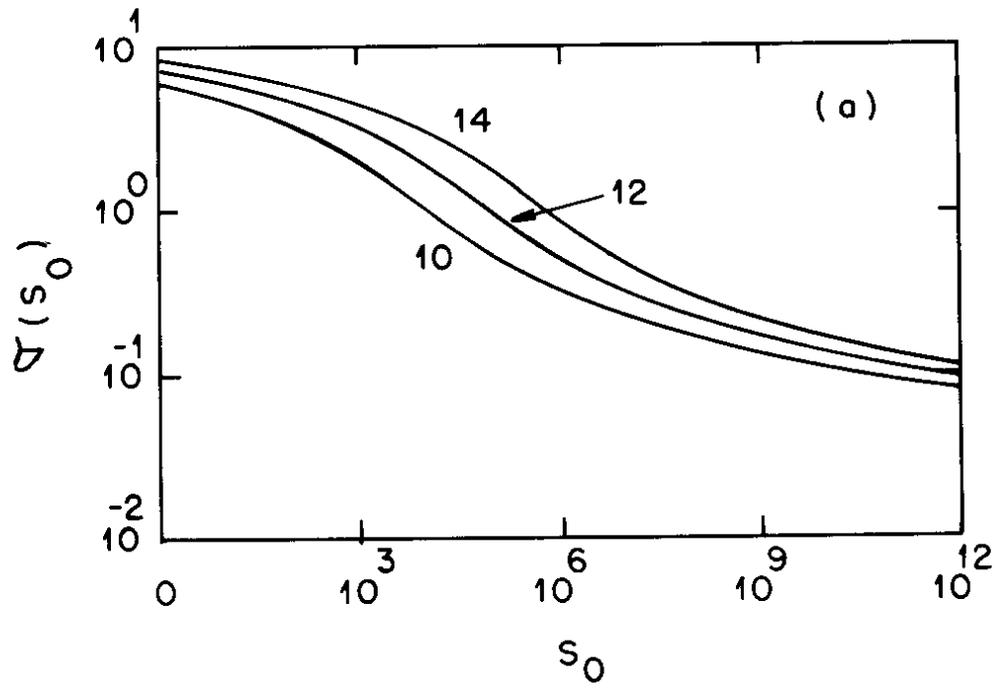
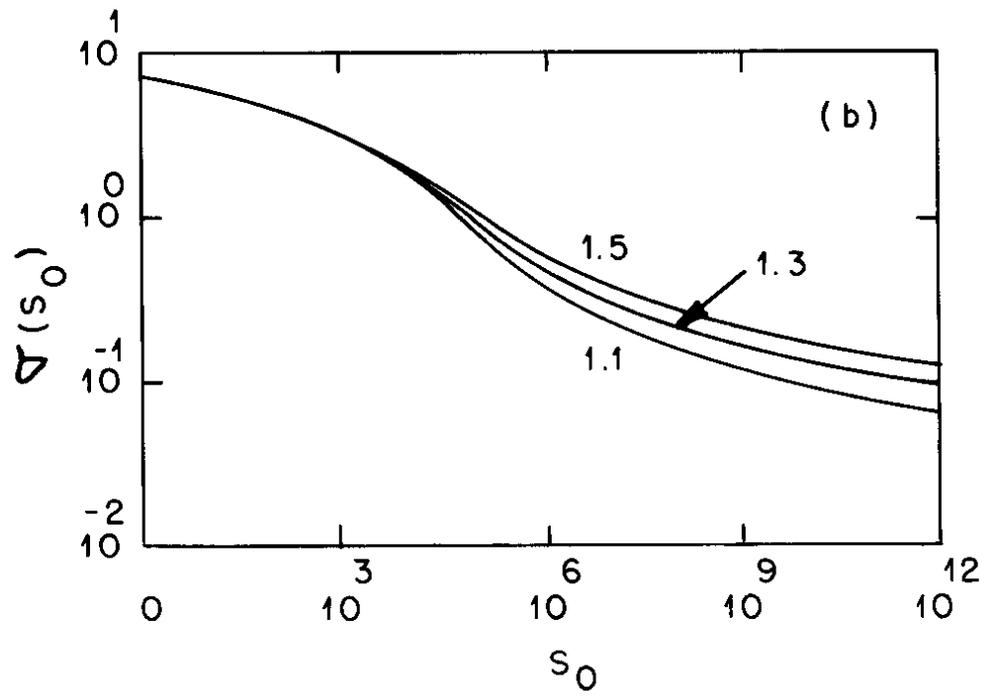
\end{document}